\documentclass[11pt,a4paper]{article}

\usepackage{mathtools,xcolor}
\usepackage{authblk} 
\usepackage{algpseudocode,algorithmicx,algorithm}

\usepackage{mathrsfs}
\usepackage{latexsym,bm,physics}
\usepackage{amsmath,amsfonts,amsmath,amssymb,amsthm}
\usepackage{extarrows}

\usepackage{graphicx,subfigure,epstopdf,float}
\usepackage{enumerate,cases,multirow}
\usepackage{makecell}
\usepackage{caption}

\usepackage{longtable,colortbl,arydshln,threeparttable}
\definecolor{mygray}{gray}{.9}

\usepackage{indentfirst}
\usepackage[top=25mm,bottom=20mm,left=25mm,right=20mm]{geometry}
\usepackage{cite}

\usepackage{listings}

\usepackage{makeidx}        
\usepackage{booktabs}
\usepackage[bookmarksnumbered,colorlinks,citecolor=red,linkcolor=red,hyperindex,linktocpage=true]{hyperref}

\newcommand{\bb}{\boldsymbol}

\def \d {\mathrm{d}}
\def \e {\mathrm{e}}
\def \i {\mathrm{i}}

\newcounter{parentalgorithm}

\makeatother

\newtheorem{theorem}{Theorem}[section]

\theoremstyle{remark}

\numberwithin{equation}{section}

\begin{document}

\title{Quantum Simulation for Partial Differential Equations with Physical Boundary or Interface Conditions}
\author[1,2]{Shi Jin\thanks{shijin-m@sjtu.edu.cn}}
\author[4]{Xiantao Li\thanks{xxl12@psu.edu}}
\author[1, 2, 3]{Nana Liu\thanks{nana.liu@quantumlah.org}}
\author[1]{Yue Yu\thanks{terenceyuyue@sjtu.edu.cn}}
\affil[1]{School of Mathematical Sciences, Institute of Natural Sciences, MOE-LSC, Shanghai Jiao Tong University, Shanghai, 200240, P. R. China}
\affil[2]{Shanghai Artificial Intelligence Laboratory, Shanghai, China}
\affil[3]{University of Michigan-Shanghai Jiao Tong University Joint Institute, Shanghai 200240, China}
\affil[4]{Department of Mathematics, The Pennsylvania State University, University Park, Pennsylvania 16802, USA}


\maketitle

\begin{abstract}
  This paper explores the feasibility of quantum simulation for partial differential equations (PDEs) with physical boundary or interface conditions. Semi-discretisation of such problems does not necessarily yield Hamiltonian dynamics and even alters the Hamiltonian structure of the dynamics when boundary and interface conditions are included. This seemingly intractable issue can be resolved by using a recently introduced Schr\"odingerisation method \cite{JLY22SchrShort, JLY22SchrLong} -- it converts any linear PDEs and ODEs with non-Hermitian dynamics to a system of Schr\"odinger equations, via the so-called warped phase transformation that maps the equation into one higher dimension.  We implement this method for several typical problems, including the linear convection equation with inflow boundary conditions and the heat equation with Dirichlet and Neumann  boundary conditions. For interface problems we study the (parabolic) Stefan problem, linear convection, and linear Liouville equations with discontinuous and even measure-valued coefficients. We perform numerical experiments to demonstrate the validity of this approach, which helps to bridge the gap between available quantum algorithms and computational models for classical and quantum dynamics with boundary and interface conditions.
\end{abstract}


\tableofcontents

\section{Introduction}

We consider the problem of quantum simulation for partial differential equations (PDEs)  with physical boundary and interface conditions. In most practical applications, one often needs to solve PDEs in a bounded domain, in which  boundary conditions should be provided  for the problem to be solvable.
Typical physical boundary conditions include Dirichlet, Neumann, and Robin (or mixed) conditions. One also encounters interface problems when the background media is  heterogeneous, for example, waves propagation across different media, heat conduction through different materials, etc.

Numerically solving PDEs becomes challenging when the space dimension is high (for example the $N$-body Schr\"odinger equation, kinetic equations such as the Boltzmann equation),  or when there are multiple time and space scales.
These problems are often too big to be solvable for classical computers, and in recent years there are increasing activities in developing quantum algorithms that use quantum computers-- yet to be developed in the future --to solve PDEs \cite{Cao2013Poisson,Berry-2014,qFEM-2016,Costa2019Wave,Engel2019qVlasov,Childs-Liu-2020,Linden2020heat,
Childs2021high,JinLiu2022nonlinear,GJL2022QuantumUQ,JLY2022multiscale}, many of which rely upon the exponential acceleration advantages in quantum linear systems of equations \cite{HHL2009,Childs2017QLSA,Costa2021QLSA,Berry-2014,BerryChilds2017ODE,Childs-Liu-2020,Subasi2019AQC}. One way to develop quantum PDE solvers is to first discretise the spatial variables to get a system of ordinary differential equations (ODEs), which in turn is solved by quantum ODE solvers \cite{Berry-2014,BerryChilds2017ODE,Childs-Liu-2020}. In particular, when the resulting ODE is also a Hamiltonian system, one can perform  quantum simulations with less time complexity than quantum ODE solvers or other quantum linear algebra solvers (e.g., the quantum difference methods \cite{Berry-2014,JLY2022multiscale}). Thus the design of quantum simulation algorithms for solving linear PDEs  become interesting and important. See a very recent proposal using block-encoding \cite{An2022blockEncodingODE}.

In a recent work, a new, simple and generic framework coined as {\it Schr\"odingerisation} was introduced \cite{JLY22SchrShort, JLY22SchrLong} that allows quantum simulation for {\it all} linear PDEs and ODEs, and even some iterative methods in linear algebra
\cite{JinLiu-LA}. The idea is to use a warped phase transform that maps the equations to one higher dimension, which, in the Fourier space, become a system of Schr\"odinger's equations! The method is extended to solve open quantum systems in a bounded domain where artificial boundary conditions --which are not unitary operators --are needed \cite{JLLY23ABC}.

In this paper, we explore the Schr\"odingerisation technique for PDEs with physical boundary and interface conditions.
These conditions do not have unitary properties thus are not naturally suitable for quantum simulations. While a (homogeneous) PDE, when spatially discretised, becomes a homogeneous system of ODEs or dynamical systems, the boundary conditions, when numerically discretised, could contribute to an inhomogeneous term in the dynamical system.  Our idea, as laid out in \cite{JLY22SchrShort, JLY22SchrLong}, is to introduce an auxiliary variable such that the extended system becomes homogeneous again. Then one can adapt the Schr\"odingerisation technique to turn them into a Schr\"odinger or unitary system, thus allowing direct quantum simulation.

Interface conditions, on the other hand, need extra attention when solved numerically. These problems are often modeled by PDEs with discontinuous coefficients. The immersed interface
methods \cite{LL94IIM, peskin2002immersed}, which incorporate the physical interface conditions into the numerical fluxes,  are among the most popular methods to numerically treat the interface conditions. In the case of wave propagation through heterogeneous media, Hamiltonian-Preserving schemes were proposed by Jin and Wen \cite{JinWen-CMS, JW06optics2}, where the transmission and reflection of waves crossing the interface are naturally built into the numerical fluxes. We then solve the discretised system based on the above methods via Schr\"odingerisation.

We choose some prototype PDEs with boundary and interface conditions to showcase these methods. Among the PDEs we study include parabolic and convection equations, with Dirichlet and Neumann boundary conditions. For the interface problems, we study the (parabolic) Stefan problem, linear convection, and linear Liouville equations with discontinuous and even measure-valued coefficients.

The paper is organized as follows. In section 2 we briefly review the Schr\"odingerisation technique for linear dynamical systems. Section 2 studies linear convection equation with inflow (Dirichet) boundary conditions.  In section 4 both Dirichlet and Neumann boundary conditions are considered. Section 5 studies linear convection and heat equations with discontinuous coefficients describing interfaces. In section 6 we study geometric optics problems across interfaces where the numerical fluxes need to take into account the partial transmissions and reflections. The paper is concluded in Section 7.

\section{Quantum simulations via Schr\"odingerisation}\label{sec:Schrodingerisation}

Of central importance in quantum computing algorithms is Hamiltonian simulation techniques. Assuming access to a Hermitian matrix $H$, they construct a quantum circuit that implements the unitary operator $U=\exp (-\i tH).$ Equivalently, it provides a route to evolve the time-dependent Schr\"odinger equation,
\begin{equation}\label{eq: tdse}
 \i \partial_t \psi = H \psi.
\end{equation}
More generally, one can simulate the unitary evolution driven by a time-dependent Hamiltonian $H(t)$ \cite{low2018hamiltonian}. The
 Schr\"odingerisation technique  \cite{JLY22SchrShort,JLY22SchrLong} extends Hamiltonian simulation methods to the solution of PDEs. The technique turns a general dynamical system into a (decoupled) system of  Schr\"odinger equations, thus paving the way to solve general time-dependent PDEs using Hamiltonian simulation techniques. In this section, we review the main steps in the Schr\"odingerisation procedure.

In practice, a PDE in a physical domain can be first discretised in space, while keeping the continuous dependence on time. A wide variety of methods are available for this purpose, including  finite difference methods, finite element elements, spectral methods, etc.  Such a spatial discretisation strategy reduces the problem to an ODE system, which can be expressed in the following general form,
\begin{equation}\label{ODElinear}
 \begin{cases}
 \frac{\d \bb{u}(t)}{\d t} = A(t) \bb{u}(t) + \bb{b}(t), \\
 \bb{u}(0) = \bb{u}_0,
 \end{cases}
\end{equation}
where $\bb{u}, \bb{b} \in \mathbb{C}^n$  and $A \in \mathbb{C}^{n\times n} $. In general, $A $ is non-Hermitian, i.e.,  $A^{\dagger} \neq A$, where "$\dagger$" denotes conjugate transpose.  We first show that it suffices to assume that $\bb{b}(t) = \bb{0}$. Otherwise one can instead consider the augmented system:
\[ \begin{cases}
 \frac{\d \bb{u}(t)}{\d t} = A \bb{u}(t) + \bb{b}(t) v,  \qquad \bb{u}(0) = \bb{u}_0,\\
 v_t = 0, \qquad v(0) = 1,
 \end{cases}\]
where the second equation gives $v(t) \equiv 1$, which leads to the original ODE system.  The above ODEs can be written in the following compact form
\begin{equation}\label{eq: general}
\begin{cases}
 \frac{\d \tilde{\bb{u}}(t)}{\d t} = \tilde{A} \tilde{\bb{u}}(t)  \\
 \tilde{\bb{u}}(0) = \tilde{\bb{u}}_0
 \end{cases},\qquad \tilde{\bb{u}} = \begin{bmatrix} \bb{u} \\ v \end{bmatrix},  \qquad\tilde{A} = \begin{bmatrix}
A & \bb{b}(t) \\
\bb{0}^T & 0
\end{bmatrix}, \qquad
\tilde{\bb{u}}_0 = \begin{bmatrix} \bb{u}_0 \\ 1 \end{bmatrix},\end{equation}
where the zero vector $\bb{0}$ has the same size as $\bb{b}$. For this reason, without loss of generality,  we assume $\bb{b} = \bb{0}$ in the following.

Now we return to the general form \eqref{ODElinear}. We begin by  decomposing $A$ into a Hermitian term and an anti-Hermitian term:
\[A = H_1 + \i H_2,\]
where
\[
H_1 = \frac{A+A^{\dagger}}{2} = H_1^{\dagger}, \qquad H_2 = \frac{A-A^{\dagger}}{2 \i} = H_2^{\dagger}.
\]

A natural assumption is that \eqref{ODElinear} inherits the stability of the original PDE, in that the eigenvalues of $A$ have non-positive real parts.  The stability property implies that $H_1$ is negative semi-definite. Using the warped phase transformation $\bb{v}(t,p) = \e^{-p} \bb{u}(t)$ for $p\ge 0$ and symmetrically extending the initial data to $p<0$, the ODEs are then transformed to a system of linear convection equations \cite{JLY22SchrShort,JLY22SchrLong}:
\begin{equation}\label{u2v}
\begin{cases}
 \frac{\d}{\d t} \bb{v}(t,p) = A \bb{v}(t,p) = - H_1 \partial_p \bb{v} + \i H_2 \bb{v}, \\
 \bb{v}(0,p) = \e^{-|p|} \bb{u}_0.
 \end{cases}
\end{equation}
Let $Q^{-1} H_1 Q = \text{diag}(\lambda_1, \cdots, \lambda_n)$ with $\lambda_j\le 0$ and define $\tilde{\bb{v}} = Q^{-1} \bb{v}$. When neglecting the imaginary part, one can find that the wave $\tilde{\bb{v}}_j$ moves from right to left with speed $s_j = |\lambda_j|$.

\begin{figure}[!htb]
  \centering
  \includegraphics[height=5cm,width=10cm]{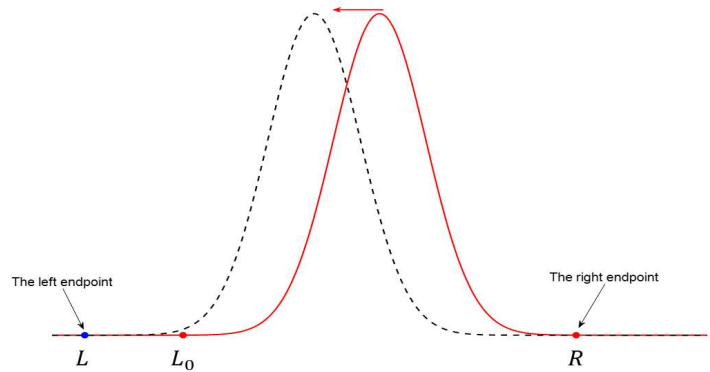}\\
  \caption{Schematic diagram for the computational domain of $p$}\label{fig:domain}
\end{figure}

For numerical implementation, it is natural and convenient to introduce  $\alpha = \alpha(p)$ in the initial data of \eqref{u2v} for $p<0$:
\begin{equation}\label{u2valpha}
\begin{cases}
 \frac{\d}{\d t} \bb{v}(t,p) = A \bb{v}(t,p) = - H_1 \partial_p \bb{v} + \i H_2 \bb{v}, \\
 \bb{v}(0,p) = \e^{-\alpha |p|} \bb{u}_0.
 \end{cases}
\end{equation}
To match the exact solution, $\alpha(p) = 1$ is necessary for the region $p> 0$. In the $p>0$-domain, we will truncation the domain at $p=R$, where $R$ is sufficiently large such that $\e^{-R} \approx 0$. We will choose a large $\alpha$ for $p<0$ so the solution (see Fig.~\ref{fig:domain}) will have a support within a relatively small domain.  Since  the wave $\tilde{\bb{v}}_j$ moves to the left, one needs to choose the artificial boundary at $p=L<0$, for $|L|$ large enough such that $\tilde{\bb{v}}_j$, initially almost compact at $[L_0, R]$, will not reach the point $p=L$ during the duration of the computation. This will allow  to use periodic boundary condition in $p$ for spectral approximation.

The solution $\bb{u}(t)$ can be restored by
\[\bb{u}(t) = \int_0^\infty \bb{v}(t,p) \d p \qquad \mbox{or} \qquad \bb{u}(t) = \e^{p_k t} \bb{v}(t,p_k), \quad {\text{ for some} } \quad p_k>0. \]
A more intuitive view is by discretising the $p$ domain and concatenating the corresponding function for each $p$. Toward this end,  we choose uniform mesh size $\Delta p = (R-L)/N_p$ for the auxiliary variable with $N_p$ being an even number, with the grid points denoted by $a = p_0<p_1<\cdots<p_{N_p} = b$. Let the vector $\bm{w}$ be the collection of the function $\bm v$ at these grid points, defined more precisely as follows,
\[\bm{w} = [\bm{w}_1; \bm{w}_2; \cdots; \bm{w}_n], \]
with ``;'' indicating the straightening of $\{\bm{w}_i\}_{i\ge 1}$ into a column vector. This can also be expressed as a superposition state by $\ket{k}$ as a new basis,
\[
 \bm{w}_i = \sum_k \bm{v}_i (t,p_k) \ket{k}.
\]
By applying the discrete Fourier transformation in the $p$  direction, one arrives at
\begin{equation}\label{heatww}
\frac{\d}{\d t} \bb{w}(t) = -\i ( H_1 \otimes P_\mu ) \bb{w} + \i (H_2 \otimes I) \bb{w}.
\end{equation}
At this point, we have successfully mapped the dynamics back to a Hamiltonian system. Here, $P_\mu$ is the matrix expression of the momentum operator $-\i\partial_p$, given by
\[P_\mu = \Phi D_\mu \Phi^{-1},  \qquad D_\mu = \text{diag}(\mu_{-N_p/2}, \cdots, \mu_{N_p/2-1}), \]
where $\mu_l = 2\pi l/(R-L)$ are the Fourier modes and
\[\Phi = (\phi_{jl})_{M\times M} = (\phi_l(x_j))_{N_p\times N_p}, \qquad \phi_l(x) = \e ^{\i \mu_l (x-L)}. \]
By a change of variables $\tilde{\bb{w}} = (I \otimes \Phi^{-1})\bb{w}$, one has
\begin{equation}\label{generalSchr}
\frac{\d}{\d t} \tilde{\bb{w}}(t) = -\i ( H_1 \otimes D_\mu ) \tilde{\bb{w}} + \i (H_2 \otimes I) \tilde{\bb{w}}.
\end{equation}
This is more amenable to an approximation by a quantum algorithm. In particular, if $H_1$ and $H_2$ are sparse, then \eqref{generalSchr} is a Schr\"odinger equation with the Hamiltonian $H = H_1 \otimes D_\mu - H_2 \otimes I$ that inherits the sparsity.

With the state vector encoding $\tilde{\bb{w}}$, one can apply the quantum Fourier transform on $p$ to get back to $\bb{w}$ and then restore $\bb{u}$ by projecting onto some basis $\ket{k}$ or computing the observable induced by the numerical integration. See \cite{JLY22SchrShort} for details on how to retrieve the quantum state with amplitudes proportional to $\bb{u}$ and subsequently the observables.

This article aims to demonstrate the feasibility of quantum simulation for PDEs with physical boundary conditions and interface conditions.
To assess the algorithm complexity associated with the implementation of the time-dependent Schr\"odinger equation, one can use the recent results by Berry et al. \cite[Theorem 10]{BerryChilds2020TimeHamiltonian}, although the other algorithms can also be used for the assessment. Here we simply highlight the query complexity,
\begin{theorem}
    The TDSE \eqref{eq: tdse} with an $s$-sparse Hamiltonian $H(t)$ can be simulated from $t=0$ to $t=T$ within error $\epsilon$ with query complexity,
    \begin{equation}
        \mathcal{O}\left( s \norm{H}_\text{max,1} \frac{\log (s \norm{H}_\text{max,1}/\epsilon )}{ \log\log(\norm{H}_\text{max,1}/\epsilon) } \right).
    \end{equation}
 \end{theorem}

Here the norm is defined as,
\begin{equation}
    \norm{H}_\text{max,1} =\int_0^T \norm{H}_\text{max}(t) dt, \quad
\norm{H}_\text{max}(t):= \max_{i,j} \abs{H_{i,j}(t)}.
\end{equation}

The strategy of implementing a semi-discrete approximation of a PDE system using Schr\"odingerisation is quite general. In the next few sections, we will illustrate how physical boundary conditions can be incorporated into this framework.

\section{Linear convection equation with inflow boundary conditions}

We first discuss how hyperbolic PDEs can be treated using the Schr\"odingerisation technique.
As a specific example, we consider the quantum simulations for solving the first-order hyperbolic equation
\begin{equation*}
    u_t + \nabla \cdot (c(x) u) = 0, \quad \bb{x} = (x_1,x_2,\cdots,x_d) \in (a,b)^d,
\end{equation*}
where $c(x) = [c_1(x),c_2(x),\cdots,c_d(x)]^T$ and $u = u(t,x_1,x_2,\cdots, x_d)$. This is a typical linear wave equation through inhomogeneous media.  It also appears in the linear representation of nonlinear dynamics, see the Liouville equation in \cite{JinLiu2022nonlinear,JLY22nonlinear} for instance.
For simplicity we set $c_1(x) = \cdots = c_d(x) \equiv 1$ in what follows and impose the inflow boundary conditions.

To construct a spatial discretisation, we  introduce $N_x+1$ spatial mesh points $0<x_{i,0}<x_{i,1}<\cdots<x_{i,N_x} = 1$ by $x_{i,j} = a+j \Delta x$ in the $x_i$-direction, where $\Delta x = (b-a)/N_x$. Let $\bb{j} = (j_1,j_2,\cdots, j_d)$. In addition, we consider the upwind scheme, which  can be written as
\begin{equation}
    \frac{\d }{\d t} u_{\bb{j}}(t) + \sum\limits_{k=1}^d \frac{u_{\bb{j}}(t)- u_{\bb{j}-\bb{e}_k}(t)}{\Delta x} = 0,
\end{equation}
where $\bb{e}_k = (0,\cdots,0,1,0,\cdots,0)$ with $k$-th entry being 1 and $j_k = 1,2,\cdots, N_x$ and we assume that the scheme is also applied to the right endpoint $x = x_{i,N_x}$ along each dimension. Denote by $\bb{u}$ the vector form of the $d$-order tensor $(u_{\bb{j}}) = (u_{j_1,j_2,\cdots, j_d})$:
\[\bb{u} = \sum\limits_{\bb{j}} u_{\bb{j}} \ket{\bb{j}}
= \sum\limits_{j_1,j_2,\cdots,j_d = 1}^{N_x} u_{j_1,j_2,\cdots,j_d} \ket{j_1,j_2,\cdots,j_d},\]
where we have used the notation in quantum computation. One can refer to \cite{JLY22nonlinear} for details. The associated linear system can be represented as
\[\sum\limits_{\bb{j}} \Big( \frac{\d }{\d t} u_{\bb{j}}(t) + \sum\limits_{k=1}^d \frac{u_{\bb{j}}(t)- u_{\bb{j}-\bb{e}_k}(t)}{\Delta x} \Big)\ket{\bb{j}} = \bb{0}.\]
Due to the wave propagation nature of the PDE, the boundary conditions should only be imposed on one side of the boundaries.
To write the above system in matrix form, let us assume that $u_{\bb{j}}$ can be decomposed as $u_{\bb{j}} = u_{j_1} u_{j_2} \cdots u_{j_d}$.
Noting that $u_{j_1, \cdots, j_d}$ ($j_k = 0$) are the given inflow boundary values in the $x_k$-direction, we have
\begin{align*}
\sum\limits_{\bb{j}} u_{\bb{j}-\bb{e}_k}\ket{\bb{j}}
& = \sum\limits_{j_i = 1, i \ne k}^{N_x} u_{j_1,\cdots,0,\cdots,j_d} \ket{j_1, \cdots, 1, \cdots, j_d} \\
& + \quad \sum\limits_{j_1 = 1}^{N_x} u_{j_1} \ket{j_1} \otimes \cdots  \otimes \sum\limits_{j_k = 2}^{N_x} u_{j_k-1}\ket{j_k} \otimes  \cdots \otimes \sum\limits_{j_d = 1}^{N_x} u_{j_d}\ket{j_d} \\
& = : \bb{u}_{0k} + \bb{u}^{(1)} \otimes \cdots \otimes T_h \bb{u}^{(k)} \otimes \cdots \otimes \bb{u}^{(d)} \\
& = \bb{u}_{0k} + ( I \otimes \cdots \otimes T_h \otimes \cdots \otimes I) (\bb{u}^{(1)} \otimes \cdots \otimes \bb{u}^{(d)})\\
& = \bb{u}_{0k} + ( I \otimes \cdots \otimes T_h \otimes \cdots \otimes I) \bb{u},
\end{align*}
where
\[\bb{u}_{0k}(t) = \sum\limits_{j_i = 1, i \ne k}^{N_x} u_{j_1,\cdots,0,\cdots,j_d} \ket{j_1, \cdots, 1, \cdots, j_d} \quad (\mbox{0 and 1 are located at the $k$-th position})\]
is the vector generated by left boundary values in the $x_k$-direction,
\[\bb{u}^{(i)} = \sum\limits_{j_i = 1}^{N_x} u_{j_i} \ket{j_i}, \qquad i = 1,2,\cdots,d\]
and
\begin{equation}\label{Th}
T_h =
\begin{bmatrix}
0   &          &           &          \\
 1   &  \ddots  &           &         \\
     &   \ddots & \ddots    &   \\
     &          &    1      & 0 \\
\end{bmatrix}_{N_x \times N_x}.\end{equation}
We therefore obtain the system \eqref{ODElinear} with
\begin{equation}\label{eq: Amat}
    A = \underbrace{L_h\otimes I \otimes \cdots \otimes I}_{d~\text{matrices}} + \cdots +
I \otimes \cdots \otimes L_h \otimes I  +  I \otimes I \otimes \cdots \otimes L_h , \quad
L_h = \frac{1}{\Delta x}(T_h-I).
\end{equation}
and
\[\bb{b}(t) = \frac{1}{\Delta x}\sum\limits_{k=1}^d \bb{u}_{0k}(t).\]

To test this approach, we consider the implementation in 1D. According to the introduction in Section \ref{sec:Schrodingerisation}, the underlying waves move from right to left with speed
\[s_j =  - \lambda_j(H_1) =  - \lambda_j \Big( \frac{A+A^T}{2} \Big) =  \frac{2}{\Delta x}  \sin^2\frac{j\pi}{2(N_x+1) }, \qquad j = 1,2,\cdots, N_x\]
when $\bb{b}(t) = \bb{0}$ is assumed.
Let $s_* = \max\{s_j\}$. We know that the fastest left moving wave will have a speed $s_*= \mathcal{O}(1/\Delta x)$. Given the evolution time $T$, we can estimate a large enough $|L|$ such that
\begin{equation}\label{tchoice}
s_* T \le L_0 - L \qquad \mbox{or} \qquad L = L_0 - s_* T.
\end{equation}
We use the backward Euler scheme for the temporal discretisation. It is important to note that the purpose of these experiments is to demonstrate that the equation \eqref{generalSchr} in the Schr\"odingerisation captures the dynamics under various physical boundary conditions. For higher dimensional problems, a quantum implementation of \eqref{generalSchr} is preferred to   classical computers, due to the less dependence of the complexity on the dimension.

\begin{figure}[!htb]
  \centering
  \includegraphics[scale=0.6]{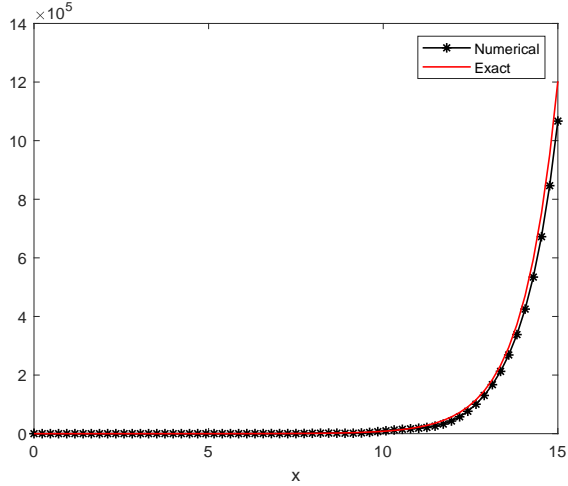}\\
  \caption{Numerical and exact solutions for the convection equation with inflow boundary condition}\label{fig:Schr_Convection_inflow}
\end{figure}

The initial and boundary values are chosen such that the exact solution is given by $u(t,x) = \e^{x-t}$. Since the speed $s_*$ scales as $\mathcal{O}(1/\Delta x)$ (we take $s_* = 2/\Delta x$), we choose a relatively large spatial domain $[a,b] = [0,10]$ such that $|L|$ is not very large. In the numerical test, we choose $T = 1$, $L_0 =-1$, $R = 10$ and $\alpha(p) = 10$ for $p<0$. For the spatial and $p$ domains, we take $N_x = N_p = 64$, then the estimated $L = -9.5333$. We also take $N_t = 100$ for the temporal discretisation. The result at $t=T$ is displayed in Fig.~\ref{fig:Schr_Convection_inflow}, from which we observe the numerical solution is well matched with the exact one.

\section{Heat equation with Dirichlet or Neumann boundary conditions}

In this section, we demonstrate how to solve parabolic PDEs with quantum simulations. Toward this end, we consider
the linear heat equation
\begin{equation*}\label{eq:heat}
\begin{cases}
& u_t(\bb{x},t) = \Delta u(\bb{x},t) \quad  \text{in}~~ \Omega :=(a,b)^d, \quad 0<t<1, \\
& u(\bb{x},0) = u_0(\bb{x}), \\
& u(\cdot, t) = 0 \quad \text{on}~~\partial \Omega,
\end{cases}
\end{equation*}
where $u_0(\bb{x})$ is the initial data. For periodic boundary conditions, one can refer to the detailed paper \cite{JLY22SchrLong} on the Schr\"odingerisation approach, where the Fourier spectral approach is used to discretise both the spatial and the auxiliary variables. For other types of boundary conditions, we consider the finite difference discretisation for the spatial domain and use the spectral discretisation for the auxiliary variable.

Let us first consider the Dirichlet boundary conditions in 1D. The central difference discretisation gives
\begin{equation}\label{heatdiscretisation}
\frac{{\rm d}}{{\rm d}t}u_j(t) = \frac{u_{j-1}(t) - 2u_j(t) + u_{j+1}(t)}{\Delta x^2},  \quad j = 1,\cdots, N_x-1.
\end{equation}
Let $\bb{u}(t) = [u_1(t), \cdots, u_{N_x-1}(t)]^T$. One gets the system \eqref{ODElinear} with
\[A = \frac{1}{\Delta x^2}
\begin{bmatrix}
-2  &  1       &           &      &    \\
 1  & -2       & \ddots    &      &    \\
    &  \ddots  & \ddots    &  \ddots    &    \\
    &          & \ddots    & -2   & 1  \\
    &          &           &  1   & -2 \\
\end{bmatrix}_{(N_x-1) \times (N_x-1)}
, \qquad
\bb{b}(t) = \frac{1}{\Delta x^2}
\begin{bmatrix}
u_0(t) \\
0  \\
\vdots\\
0 \\
u_{N_x}(t) \\
\end{bmatrix}.
\]
For $d$ dimensions, the solution vector is defined by
\[\bb{u}_{h,d} = \sum\limits_{j_1,\cdots,j_d = 1}^{N_x-1} u_{j_1,\cdots,j_d} \ket{j_1,\cdots,j_d}. \]
The corresponding coefficient matrix and right-hand vector in $d$ dimensions will be replaced by
\begin{equation}\label{heatmatrixd}
A_{h,d} = \underbrace{A\otimes I \otimes \cdots \otimes I}_{d~\text{matrices}} + I \otimes A\otimes \cdots \otimes I + \cdots + I \otimes I \otimes \cdots \otimes A
\end{equation}
and
\[\bb{b}_{h,d}(t) = \frac{1}{\Delta x^2} \sum\limits_{k=1}^d ( \bb{u}_{0,k} + \bb{u}_{N_x,k} ),\]
where
\begin{align*}
& \bb{u}_{0,k} = \sum\limits_{j_i = 1, i \ne k}^{N_x-1} u_{j_1,\cdots,0,\cdots,j_d} \ket{j_1, \cdots, 1, \cdots, j_d}, \\
& \bb{u}_{N_x,k} = \sum\limits_{j_i = 1, i \ne k}^{N_x-1} u_{j_1,\cdots,N_x,\cdots,j_d} \ket{j_1, \cdots, N_x-1, \cdots, j_d}.
\end{align*}

\medskip

Here we present a numerical test. In the implementation, the initial and boundary values are chosen such that the exact solution is given by $u(t,x) = \e^{-\pi^2 t} \sin(\pi x)$.
As analyzed in the previous section, the associated waves move from right to left with speed
\[s_j =  - \lambda_j \Big( \frac{A+A^T}{2} \Big) =  \frac{4}{\Delta x^2}  \sin^2\frac{j\pi}{2N_x }, \qquad j = 1,2,\cdots, N_x-1\]
when $\bb{b}(t) = \bb{0}$ is assumed. The fastest speed can be chosen as $s_* = 4/\Delta x^2$. To reduce the computational cost in the $p$-direction on a classical computer, as discussed in the preceding section,  we take a relatively large spatial domain $[a,b]=[0,10]$.

\begin{figure}[!htb]
  \centering
  \includegraphics[scale=0.6]{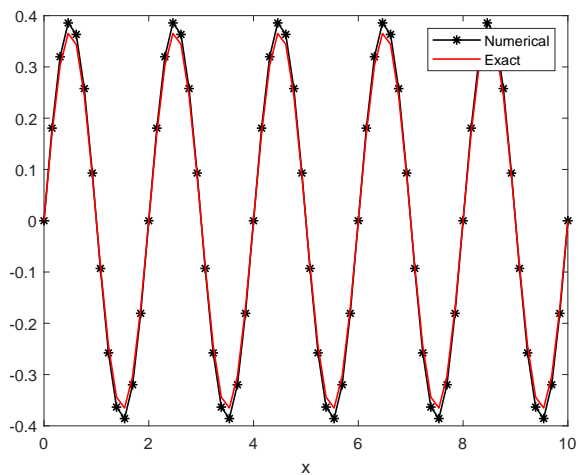}\\
  \caption{Numerical and exact solutions for the heat equation with Dirichlet boundary conditions}\label{fig:Schr_Heat_Dirichlet}
\end{figure}

Considering the exponentially decreasing factor $\e^{-\pi^2 t}$ in the exact solution, we choose the evolution time $T = 1/\pi^2$. We also choose $N_x-1 = N_p = 64$.  Other parameters are the same as for the convection equation. The estimated $L=-18.1233$. The result at $t = T$ is shown in Fig.~\ref{fig:Schr_Heat_Dirichlet}.

We next consider the heat equation with the mixed boundary conditions:
\[u(t,a) = g(t), \qquad u_x(t,b) = h(t). \]
The discretisation at the interior node is still given by \eqref{heatdiscretisation}. For the right boundary, we introduce a ghost point $x_{N_x+1} = x_{N_x} + \Delta x$ and use the central difference to discretise the first-order derivative,
\[\frac{u_{N_x+1}(t) - u_{N_x-1}(t)}{2 \Delta x} = h(t).\]
To get a closed system, we assume the discretisation in \eqref{heatdiscretisation} is valid at $x = x_{N_x}$:
\[\frac{{\rm d}}{{\rm d}t}u_{N_x}(t) = \frac{u_{N_x-1}(t) - 2u_{N_x}(t) + u_{N_x+1}(t)}{\Delta x^2}.\]
Eliminating the ghost values to get
\[\frac{{\rm d}}{{\rm d}t}u_{N_x}(t) = \frac{2 u_{N_x-1}(t) - 2u_{N_x}(t) }{\Delta x^2} + \frac{2h(t) }{\Delta x},  \quad j = N_x.\]
Let $\bb{u}(t) = [u_1(t), u_2(t), \cdots, u_{N_x}(t)]^T$. Then one gets the system \eqref{ODElinear} with
\[A = \frac{1}{\Delta x^2}
\begin{bmatrix}
-2  &  1       &           &          &                 \\
 1  & -2       & \ddots    &          &                 \\
    &  \ddots  & \ddots    &  \ddots  &               \\
    &           &  1       & -2       &   1     \\
    &           &          & 2       & -2      \\
\end{bmatrix}_{N_x \times N_x}
, \qquad
\bb{b}(t) = \frac{1}{\Delta x^2}
\begin{bmatrix}
g(t) \\
0  \\
\vdots\\
0 \\
2 h(t) \Delta x
\end{bmatrix}.
\]
We remark that the coefficient matrix in $d$ dimensions still has the form in \eqref{heatmatrixd}, with $A$ replaced by the one given here. The right-hand vector can be deduced in a similar way.

\begin{figure}[!htb]
  \centering
  \includegraphics[scale=0.6]{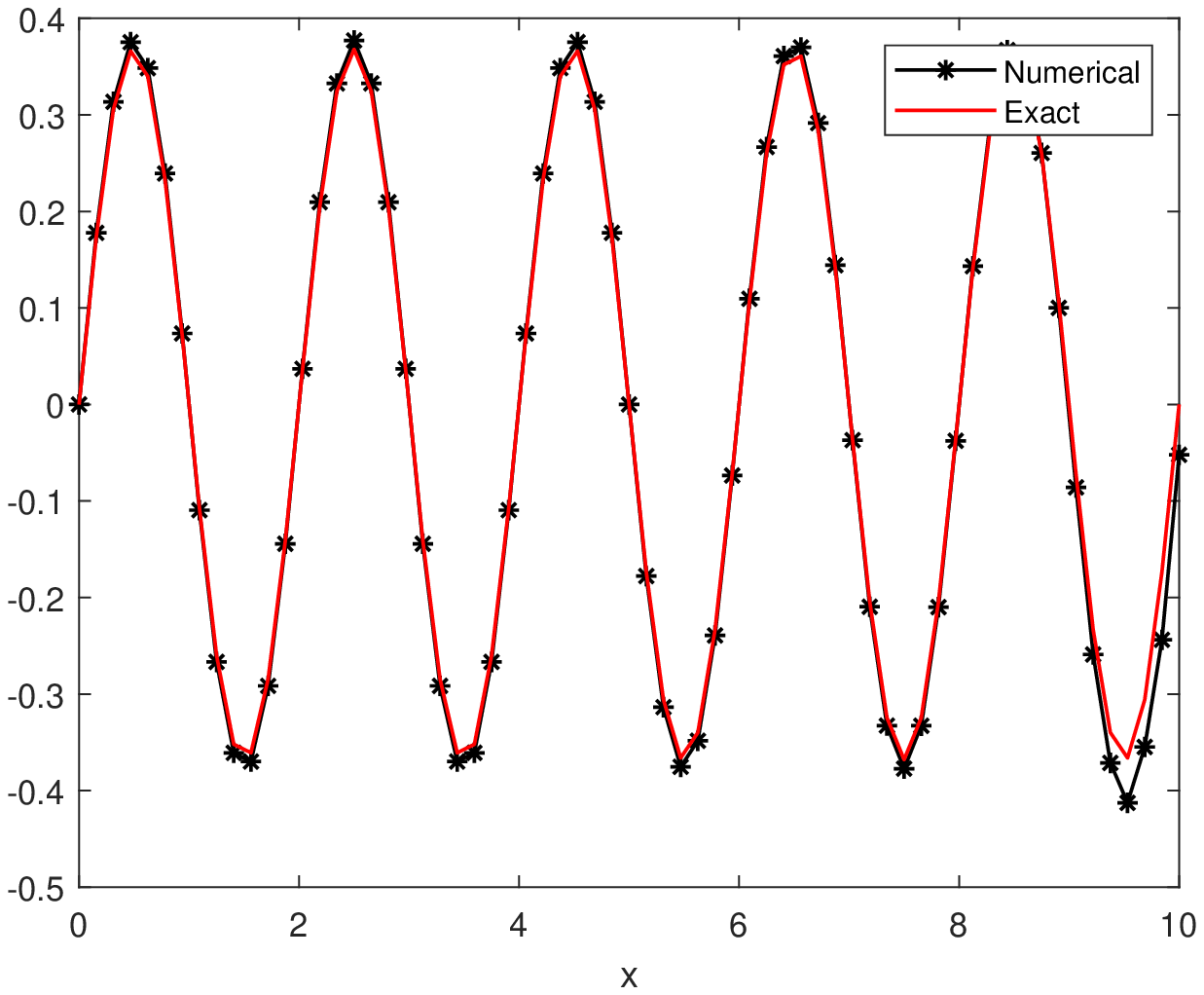}\\
  \caption{Numerical and exact solutions for the heat equation with mixed boundary conditions}\label{fig:Schr_Heat_mixed}
\end{figure}

We implement \eqref{generalSchr} on the same test problem by repeating the procedure for the case of Dirichlet boundary conditions. The snapshot is depicted in Fig.~\ref{fig:Schr_Heat_mixed}, where we set $N_x = 64$ and $N_p = 512$. One can see a good agreement between the exact and the numerical solutions.

\section{Linear PDEs for interface problems}

This section is devoted to the quantum simulation of interface problems with a fixed or moving interface.

\subsection{The linear advection equation}

We are concerned with a hyperbolic equation with discontinuous  coefficients~---~a simple interface problem in the following form:
\begin{equation}\label{interfacePro}
\begin{cases}
\partial_t u + \nabla\cdot (c(x) u) = 0, \quad t>0, x\in (-a,a)^d, \\
u(0,x) = u_0(x), \qquad x\in [-a,a]^d,
\end{cases}
\end{equation}
where $a>0$ is a constant and $c(x) = [c_1(x),\cdots,c_d(x)]^T$ is a vector for fixed $x = (x_1,\cdots,x_d)$. We assume that $c(x)$ is piecewise constant in the $x_1$-direction:
\[c_i(x) = \begin{cases}
c^->0, \qquad x_1<0 \\
c^+>0, \qquad x_1>0
\end{cases}, \qquad i = 1,\cdots,d.\]
The above equation arises in modeling wave propagation through interfaces with jumps in $c(x)$ corresponding to interfaces between different media.  For such a problem an interface condition is needed at $x_1 = 0$:
\begin{equation}\label{interfaceCond}
u(t,x^+) = \rho u(t, x^-),
\end{equation}
where $x^\pm$ represents the right and left limits in the $x_1$-direction, $\rho = 1$ corresponds to the continuity of mass $u$ or $\rho = c^-/c^+$ for the continuity of flux $cu$ \cite{Jin09hyperbolic,WJ08Immerse}.

In the following, we only consider the 1-D case. The exact solution of \eqref{interfacePro} with the interface condition \eqref{interfaceCond} can be constructed following characteristics \cite{WJ08Immerse}, given by
\[u(t,x; u_0) = \begin{cases}
u_0(x-c^- t), \qquad & x<0, \\
\rho u_0( \frac{c^-}{c^+} x - c^- t), \qquad & 0<x<c^+ t, \\
u_0(x-c^+ t), \qquad & x>c^+ t.
\end{cases}\]
Since the wave moves from left to right, we impose the boundary condition on the left endpoint.

When numerically solving \eqref{interfacePro}, the most natural approach is to build the interface condition \eqref{interfaceCond} into the numerical flux, as was  proposed in \cite{WJ08Immerse}.
Let the uniform spatial mesh be $x_j$, where $j = -N_x, \cdots, -1, 0, 1,\cdots, N_x$ and $\Delta x = a/N_x$ is the mesh size. Since $c^\pm >0$, one can apply the upwind scheme for $x\le 0$ and $x> \Delta x$:
\[\begin{cases}
\dfrac{\d u_j}{\d t} = - c^- \dfrac{u_j(t) - u_{j-1}(t) }{\Delta x}, \qquad j = -(N_x-1), \cdots, 0, \\
\dfrac{\d u_j}{\d t} = - c^+ \dfrac{u_j(t) - u_{j-1}(t) }{\Delta x}, \qquad j = 2, \cdots, N_x. \\
\end{cases}\]
Note that in the above scheme $u_0$ is the left limit of $u$ at the interface. For $j = 1$, the continuity of $cu$ gives
\[\dfrac{\d u_1}{\d t} = - \dfrac{c^+ u_1(t) - c^- u_0(t) }{\Delta x}, \qquad j = 1.\]
Let $\bb{u}(t) = [u_{-(N_x-1)}(t), \cdots, u_{N_x}(t)]^T$. One can collect the above equations as the system \eqref{ODElinear} with
\[A = \frac{1}{\Delta x} \left[\begin{array}{cccc:cccc}
-c^-  &        &  &  &   &   &   &  \\
c^- &   -c^-  &  &  &   &   &   &  \\
     &  \ddots    & \ddots &  &   &   &   &  \\
     &            & c^{-} & -c^- &   &   &   &  \\
     \hdashline
     &            & & c^{-} &  -c^+ &   &   &  \\
     &            &         &        & c^+  & -c^+  &   &  \\
     &            &         &        &       & \ddots  & \ddots  &  \\
     &            &         &        &       &         & c^+   &  -c^+\\
\end{array}\right]
\]
and  $\bb{b}(t) = [ c^-u(t,-1)/\Delta x, 0, \cdots, 0]^T$.

\begin{figure}[!htb]
  \centering
  \subfigure[$u(0.1,x)$]{\includegraphics[scale=0.35]{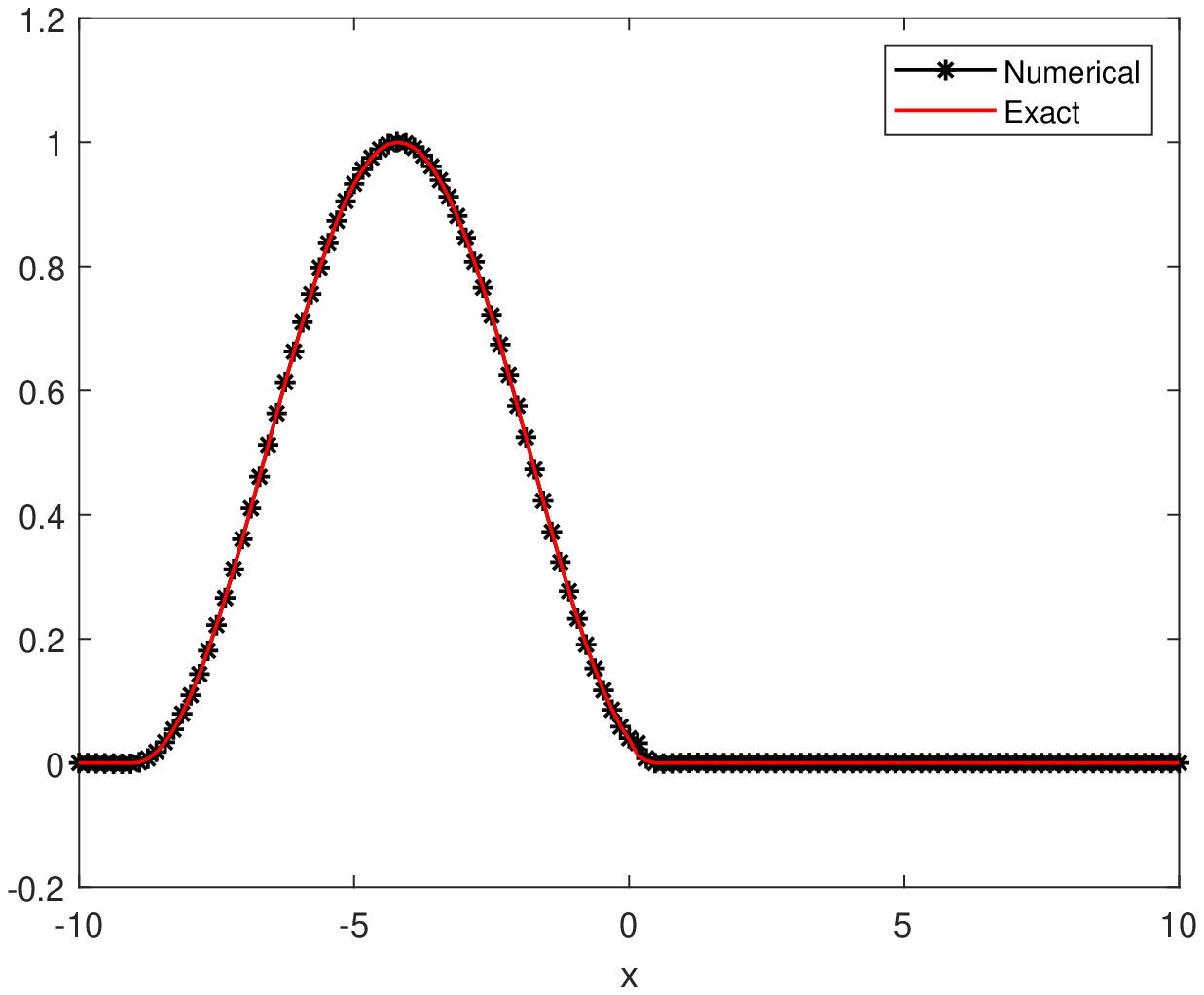}}
  \subfigure[$u(0.5,x)$]{\includegraphics[scale=0.35]{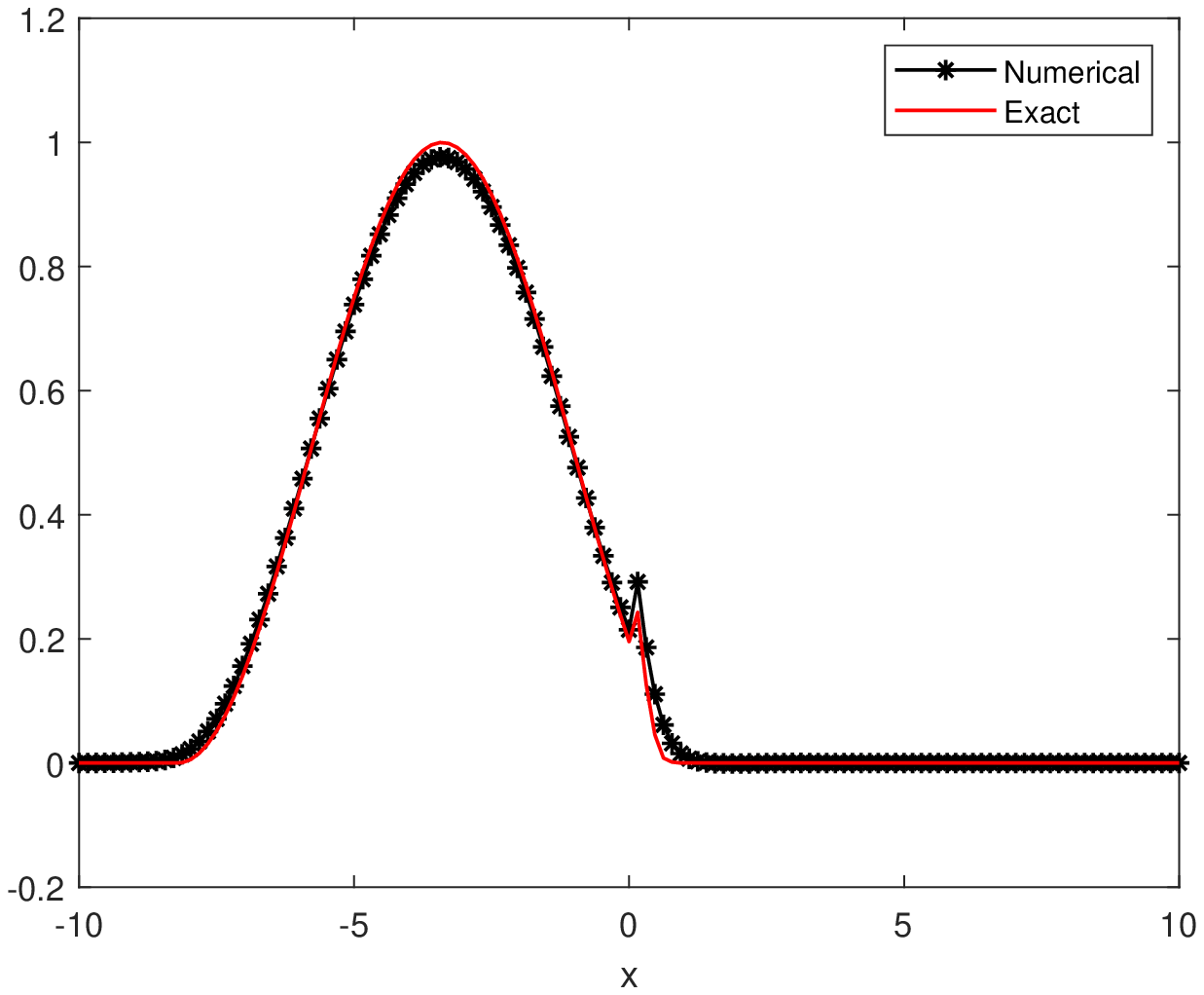}}
  \subfigure[$u(1.0,x)$]{\includegraphics[scale=0.35]{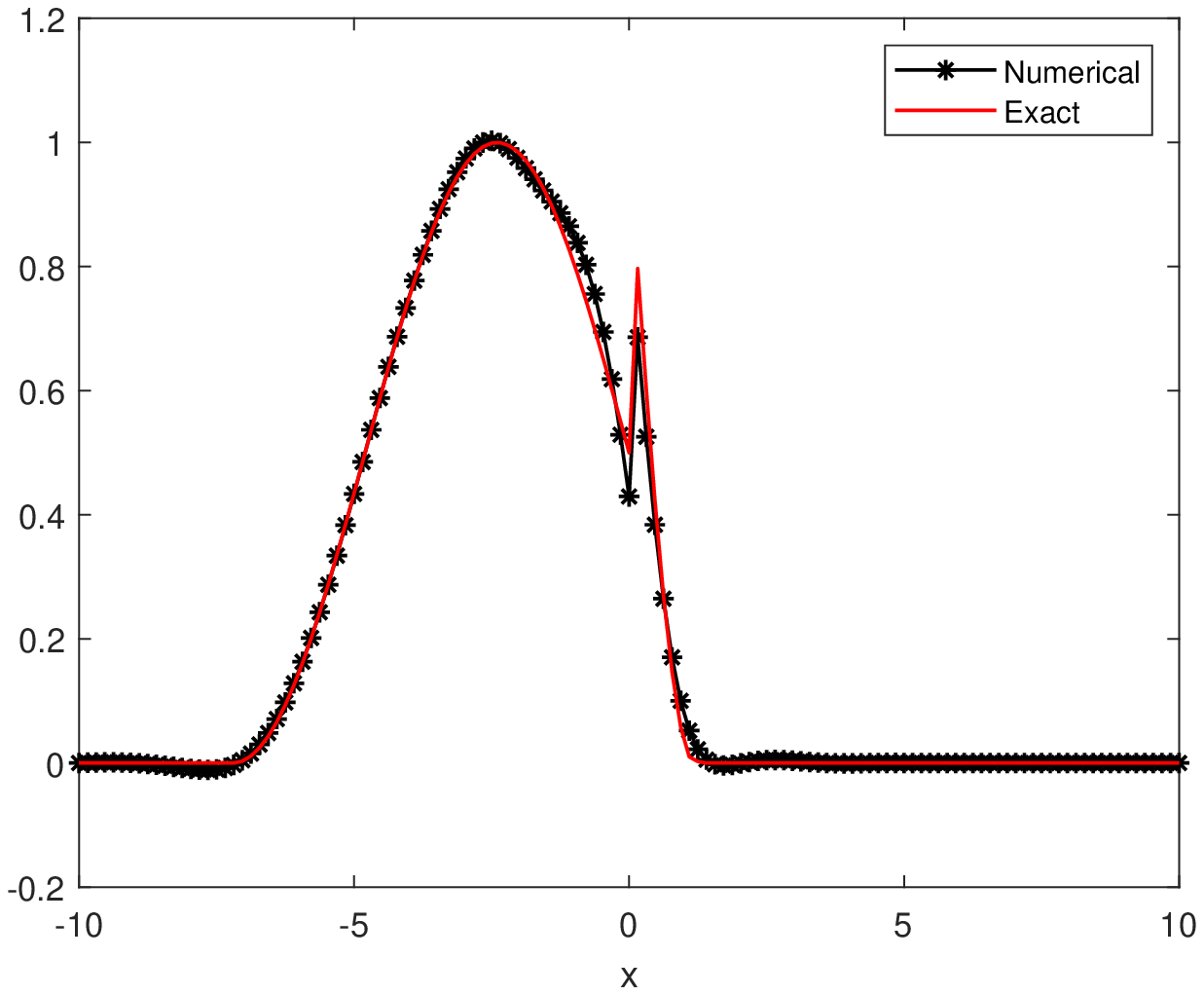}}
  \subfigure[$c(x)u(0.1,x)$]{\includegraphics[scale=0.35]{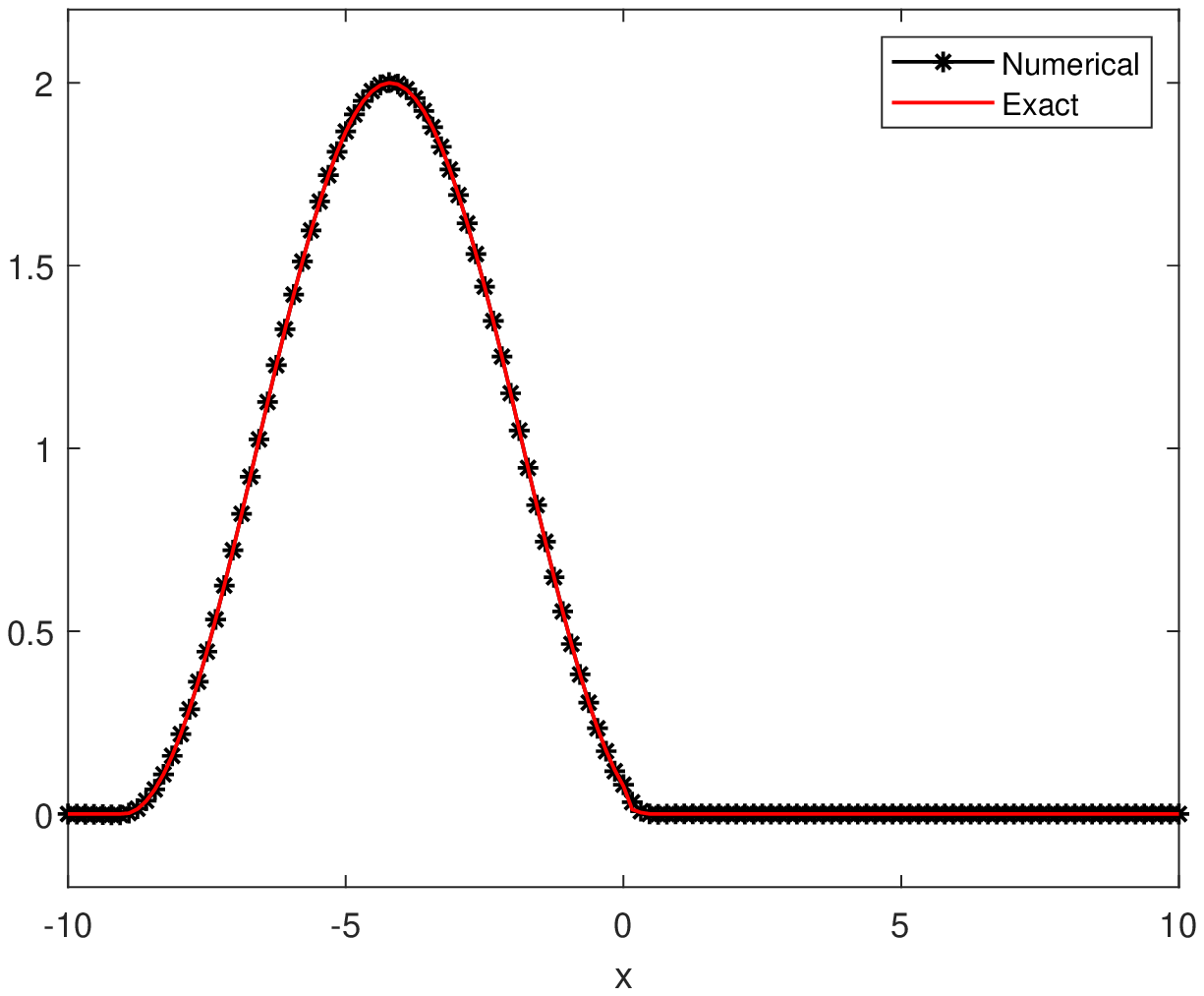}}
  \subfigure[$c(x)u(0.5,x)$]{\includegraphics[scale=0.35]{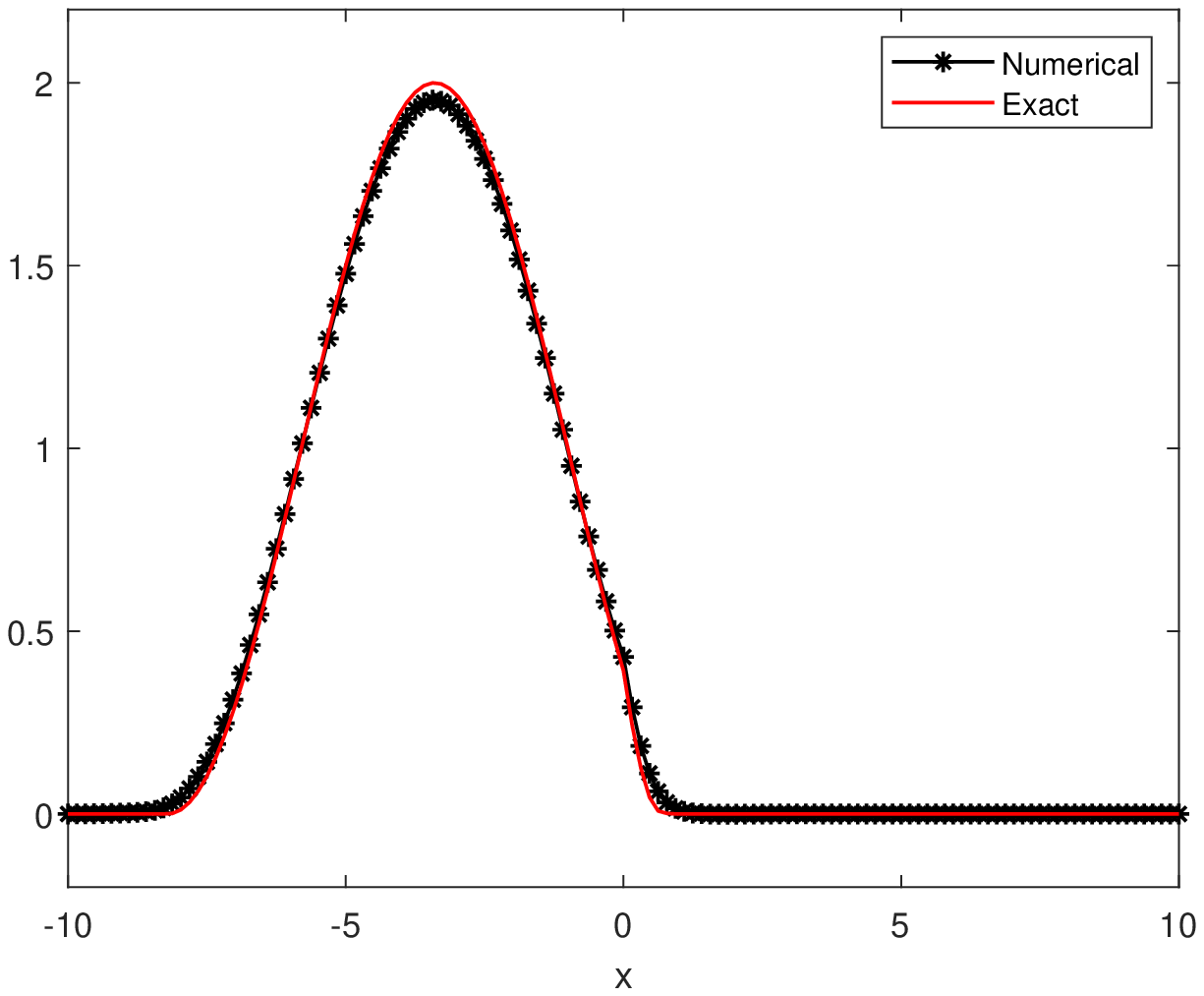}}
  \subfigure[$c(x)u(1.0,x)$]{\includegraphics[scale=0.35]{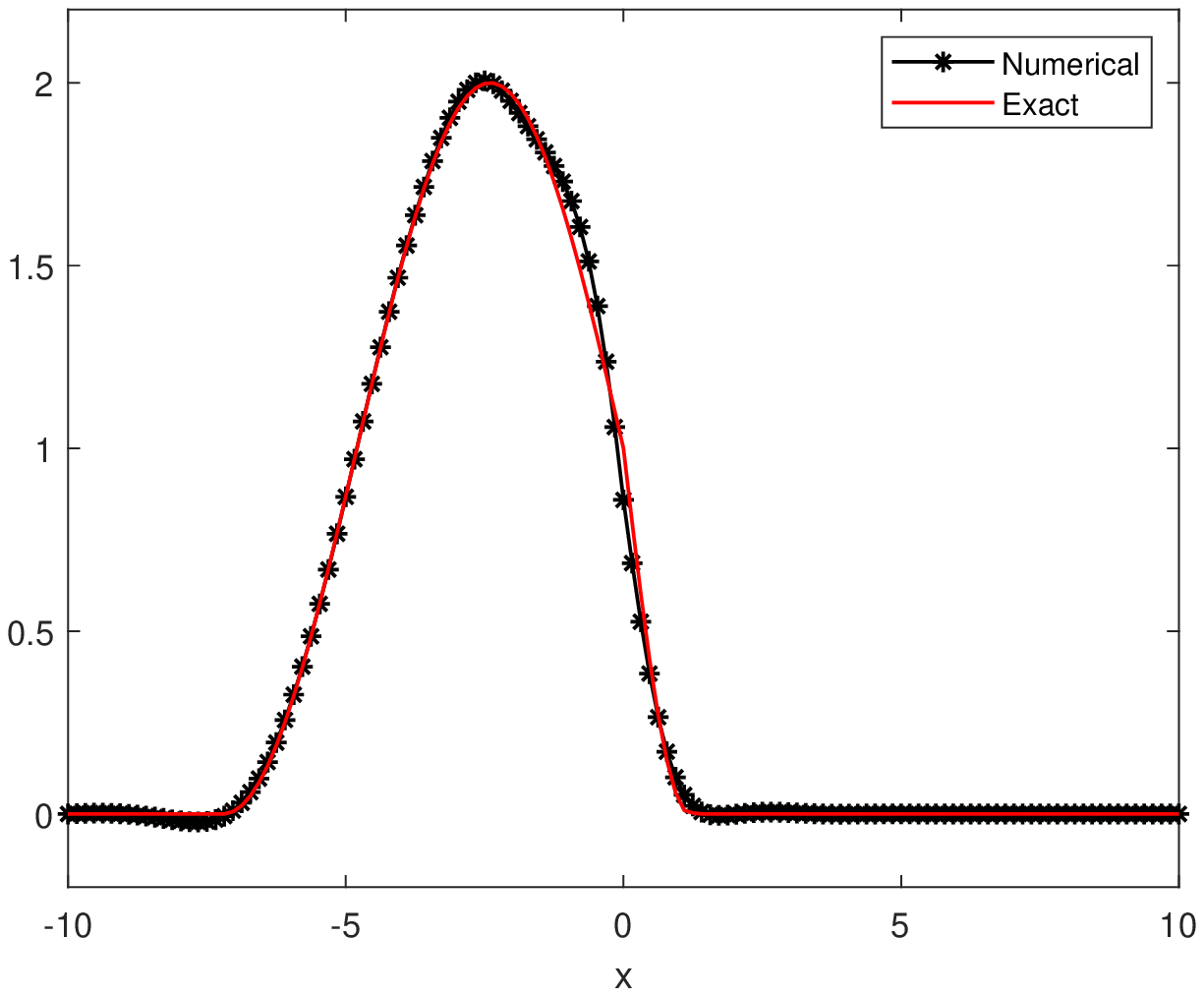}}\\
  \caption{Numerical and exact solutions for the interface problem. Top: the solution $u(T,x)$; Bottom: the flux $c(x)u(T,x)$.}\label{fig:Schr_interface}
\end{figure}

A numerical test is conducted with initial data considered in \cite{ZL97IIMadvection}:
\[u(0,x) = \begin{cases}
\dfrac12\Big(  1 + \cos  \dfrac{(x-0.28) \pi}{0.24} \Big), \qquad & -0.04\le x \le 0.52, \\
0, \qquad & \mbox{otherwise},
\end{cases}\]
which will be scaled from $(0,1)$ to $(-a,a)$ and gives the homogeneous inflow boundary condition. As in the previous sections, we choose $a = 10$ to avoid the large domain along $p$ direction.
The wave speeds are $c^- = 2$ for $x<0$ and $c^+ = 1$ for $x>0$. The fastest left-moving wave for the Schr\"odingerisation equation can be chosen as $s_*= 2 \max(c^\pm) /\Delta x$.
The numerical results of $u(t=T,x)$ are shown in the top row in Fig.~\ref{fig:Schr_interface} for the duration time $T = 0.1, 0.5$ and $1.0$, respectively. It can be seen that the Schr\"odingerisation approach captures the behaviour at the interface $x=0$, where the jump is caused by the discontinuity of $c(x)$. We also plot $c(x)u(t=T,x)$ in the bottom row in Fig.~\ref{fig:Schr_interface}, from which we clearly observe the continuity  of flux $cu$, as defined by \eqref{interfaceCond}.

\subsection{The Stefan problem}

Let $\Omega$ be an open and bounded domain in $\mathbb{R}^d$, and $\Gamma$ be a continuous interface embedded in $\Omega$. We consider the case where the interface $\Gamma: = \Gamma(t)$ varies in time. Such a problem appears in many applications, for instance, the Stefan problem for simulating temperature distribution undergoing a phase transition, where the flux jump is proportional to the velocity of the moving front \cite{rubinshteuin1971stefan}. The interface separates the domain into disjoint regions $\Omega^+$ and $\Omega^-$.  We consider quantum simulations for solving the following parabolic interface problem:
\[
u_t = \nabla\cdot (\beta(t,x) \nabla u) + f(t,x), \qquad x \in \Omega \backslash \Gamma(t),
\]
with prescribed jump conditions across the interface:
\begin{align*}
& [u]_{\Gamma} = u^+ - u^- = q_0,\\
& [\beta u_n]|_{\Gamma} = \beta^+ u_n^+ - \beta_n^- u^- = q_1,
\end{align*}
where $u_n$ denotes the normal derivative $\nabla u \cdot n$ with $n$ being a unit norm direction of the interface.
The problem in 1-D can be reformulated as
\[u_t = (\beta u_x)_x + f, \qquad x \in \Omega = (a,b),\]
with the jump condition on a moving interface $\alpha(t) \in (a,b)$:
\begin{align*}
& [u]|_{x = \alpha(t)} = u^+ - u^- = 0,\\
& [\beta u_x]|_{x = \alpha(t)} = \beta^+ u^+ - \beta^- u^- = 0.
\end{align*}
The interface separates $\Omega$ into the left and right subdomains $\Omega^-(t)$ and $\Omega^+(t)$. For simplicity, we assume $\beta(t,x)$ is a piecewise constant function:
\[\beta(t,x) = \begin{cases}
\beta^->0, \qquad & x \in \Omega^-(t), \\
\beta^+>0, \qquad & x \in \Omega^+(t).
\end{cases}\]

In the following, we assume that $t$ is fixed and simply write $\alpha(t)$ as $\alpha$. In this case one can apply the immersed interface method in \cite{LL94IIM} for the spatial discretisation. Let the uniform grid in the interval $[a,b]$ be
\[a = x_0 < x_1 < \cdots < x_N = b, \qquad  x_i = a + i h, \quad h = (b-a)/N.\]
The goal is to develop semi-discrete finite difference equations of the form
\[\frac{\d}{\d t} u_i(t)  = \gamma_{i,1}(t)u_{i-1}(t) + \gamma_{i,2}(t) u_i(t) + \gamma_{i,3}(t) u_{i+1}(t) + f_i, \qquad i = 1,2,\cdots, N-1,\]
with second-order accurate approximation to $u$ at the uniform grid points. Note that $\gamma_{i,1}, \gamma_{i,2}$ and $\gamma_{i,3}$ depend on the time variable because of the moving interface.

If we impose the Dirichlet boundary conditions, then the semi-discrete system can be written as
\[\frac{\d }{\d t} \bb{u}(t) = A(t) \bb{u}(t) + \bb{b}(t),\]
where $\bb{u}(t) = [u_1(t), \cdots, u_{N-1}(t)]^T$, and
\[A(t) = \begin{bmatrix}
\gamma_{1,2}    & \gamma_{1,3}   &                      &    \\
\gamma_{2,1}    & \ddots   &  \ddots              &    \\
                & \ddots         & \ddots               & \gamma_{N-2,3} \\
                &                & \gamma_{N-1,1}       & \gamma_{N-1,2}
\end{bmatrix}, \qquad \bb{b}(t) = \begin{bmatrix}
\gamma_{1,1}u_0(t) + f_1 \\
 f_2  \\
\vdots \\
f_{N-2} \\
\gamma_{N-1,3} u_N(t) + f_{N-1}
\end{bmatrix}.\]

Let $\alpha$ fall between $x_k$ and $x_{k+1}$, i.e., $x_k \le \alpha < x_{k+1}$ (note that $k$ depends on $t$). For $i \neq k, k+1$ the solution can be viewed as a smooth function in $[x_k,x_{k+1}]$ and one can use the standard approximation
\[\frac{1}{\Delta x^2} \Big(\beta_{i+1/2} (u_{i+1}-u_i) - \beta_{i-1/2}(u_i-u_{i-1}) \Big)\]
for $(\beta u_x)_x$ at $x = x_i$, where $\beta_{i \pm 1/2} = \beta(x_{i \pm 1/2})$. In this case one has
\[\gamma_{i,1} = \beta_{i-1/2}/h^2, \quad \gamma_{i,2} = -(\beta_{i-1/2} + \beta_{i+1/2})/h^2, \quad \gamma_{i,3} = \beta_{i+1/2}/h^2.\]
For $i=k,k+1$, following \cite{LL94IIM}, we  take
\begin{align*}
& \gamma_{k,1} = (\beta^--[\beta](x_k-\alpha)/h) / D_k, &\qquad & \gamma_{k+1,1} = \beta^-/D_{k+1},\\
& \gamma_{k,2} = (-2\beta^- + [\beta] (x_{k-1}-\alpha)/h )/D_k, & \qquad & \gamma_{k+1,2} = (-2\beta^+ + [\beta](x_{k+2}-\alpha)/h)/D_{k+1},\\
& \gamma_{k,3} = \beta^+/D_k,&
\qquad & \gamma_{k+1,3} = ( \beta^+ - [\beta](x_{k+1}-\alpha)/h)/D_{k+1},
\end{align*}
where
\[D_k = h^2 + [\beta](x_{k-1}-\alpha)(x_k-\alpha)/2\beta^-, \qquad
D_{k+1} = h^2 - [\beta](x_{k+2}-\alpha)(x_{k+1}-\alpha)/2\beta^+.\]
It is easy to show that $D_k$ and $D_{k+1}$ are positive when $\beta>0$.

\begin{figure}[!htb]
  \centering
  \includegraphics[scale=0.6]{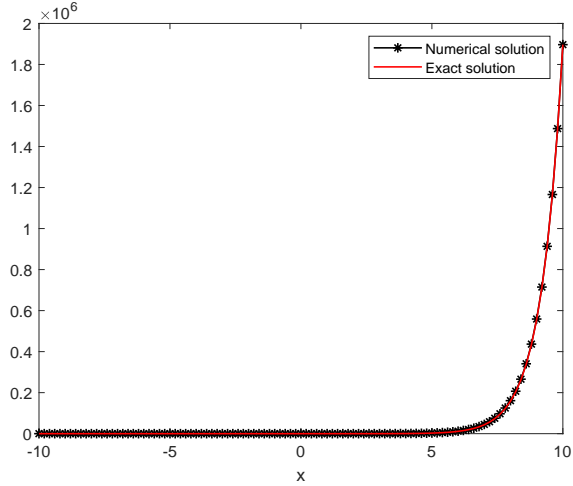}\\
  \caption{Numerical and exact solutions for the Stefan problem}\label{fig:Stefan}
\end{figure}

We implemented this model with the initial and boundary value functions  chosen such that the exact solution is
\[u(t,x) = \begin{cases}
\Big( (x-\alpha(t))^2 + \frac{1}{\beta^-} \Big)\e^x, \qquad & x \in \Omega^-(t), \\
\Big( (x-\alpha(t))^2 + \frac{1}{\beta^+} \Big)\e^x + \Big(\frac{1}{\beta^-} -\frac{1}{\beta^+}\Big)\e^{\alpha(t)} , \qquad & x \in \Omega^+(t),
\end{cases}.\]
We set $\alpha(t) = \frac12 t + \frac14$ and $\beta^- = 1$ and $\beta^+ = 2$. The backward Euler method is used for the temporal discretisation. The spatial domain is taken as $[0,10]$. We set $N = 100$ and $N_t = 100$ and display the solutions at $t = 1$ in Fig.~\ref{fig:Stefan}, from which one can see  that the Schr\"odingerisation approach gives the desired solution for the problem with interface varying in time.

\section{Geometric optics problems with partial transmissions and reflections}

In this section we are concerned with the quantum simulation of geometric optics problems when both transmissions and reflections occur at the interface.

\subsection{The geometric optics problem}

\subsubsection{The Hamilton-Jacobi equation for the geometric optics}

We consider   the linear scalar wave equation in the high frequency regime,
\begin{equation}\label{highwave}
w_{tt} - c(x)^2 \Delta w = 0,  \qquad x\in \mathbb{R}^d,
\end{equation}
where $c(x)$ is the local speed of wave propagation of the medium, or the reciprocal of the index of refraction. When the waves are of high frequency, \eqref{highwave} is a multiscale problem, where the small scale is given by the wavelength over, for example, the overall size of the computational domain.  For sufficiently high frequencies, direct numerical simulation is no longer feasible. Numerical methods based on approximations of \eqref{highwave} are needed.

The derivation of the geometrical optics equations in the linear case follows if one assumes a series expansion of the form
\begin{equation}\label{seriesExpansion}
w(t,x) = \e^{\i \omega S(t,x)} \sum\limits_{k=0}^\infty A_k(t,x) (\i \omega)^{-k}.
\end{equation}
Plugging this expression into \eqref{highwave} and collecting terms of the same order in $\omega$, one obtains separate equations for the unknown dependent variables in \eqref{seriesExpansion}. The $\mathcal{O}(\omega^2)$ terms give the equation for the phase function $S$, which satisfies the Hamilton-Jacobi-type eikonal equation \cite{ER03high}
\begin{equation}\label{HJEgeo}
\partial_t \phi + c(x) | \nabla S| = 0.
\end{equation}
Hamilton-Jacobi equations (HJE) take the following general form
\begin{align}\label{H-J}
& \partial_t S + H(\nabla S, x) = 0,\\
& S(0,x)=S_0(x)\nonumber
\end{align}
with  $t\in \mathbb{R}^+$,  $x \in \mathbb{R}^d$, $S(t,x)\in \mathbb{R}$. For the geometric optics equation \eqref{HJEgeo}, the associated Hamiltonian is given by
\begin{equation}\label{HamiltonianGeo}
H(\xi,x) = c(x)|\xi|.
\end{equation}

\subsubsection{The Liouville representation for the Hamilton-Jacobi equation}

Define $u=\nabla S \in \mathbb{R}^d$. Then $u$ solves a hyperbolic system of conservation laws in gradient form:
\begin{align}\label{forced-Burgers}
 & \partial_t u + \nabla H(u, x) = 0,\\
 &u(0,x)=\nabla S_0(x).\nonumber
\end{align}
Ref.~\cite{JinLiu2022nonlinear} constructed quantum algorithms to compute physical observables of this nonlinear problem, which is based on an exact mapping between nonlinear and linear PDEs using the level set method \cite{JinOsher2003levelset}. This approach is referred to as the linear representation approach, and it is based on an {\it exact} map from a nonlinear PDE to a linear one thus {\it no} physical information is lost, while other approaches are based on linear approximations that use {\it truncation} to linearize the problem so they are {\it not} the same physical problem as the original nonlinear one.   A more comprehensive discussion can be found in \cite{JLY22nonlinear}, where the finite difference and spectral discretisations are discussed with periodic boundary conditions applied.

We follow the linear representation approach in \cite{JinLiu2022nonlinear}. The level set function $\phi_i(t,x,p)$ can be defined by
\begin{align}\label{LS-def}
\phi_i(t, x,\xi=u(t,x))=0, \nonumber
\end{align}
where $i=1, \cdots, d$ and $\, x, \xi\in \mathbb{R}^d$, and $u(t,x)$ is the solution of Eq.~\eqref{forced-Burgers}. The \textit{zero level set} of $\phi$ is the set  $\{(t,x,p)|\phi_i(t,x,\xi)=0\}$. Since $u(t,x)$ solves Eq.~\eqref{forced-Burgers}, one can show that $\phi=(\phi_1, \cdots, \phi_d)\in \mathbb{R}^d$ solves a (linear!) Liouville equation \cite{JinOsher2003levelset}
\begin{equation} \label{LS-Liouville}
\partial_t \phi + \nabla_\xi H \cdot \nabla_x \phi - \nabla_x H \cdot \nabla_\xi \phi=0.
\end{equation}
 The initial data can be chosen as
 \begin{equation}\label{Liou-IC}
 \phi_i(0, x,\xi)=\xi_i - u_i(0,x), \quad i=1, \cdots, d.
 \end{equation}
 Then $u$ can be recovered from the intersection of the zero level
sets of $\phi_i \, (i=1, \cdots, d)$, namely
\[u(t,x)=\{\xi(t,x)| \,\phi_i(t,x,\xi)=0, \, i=1,\cdots, d \}.\]

 To retrieve physical observables (and to avoid finding the zero level set of $\phi$ which is challenging) later, \cite{JinLiu2022nonlinear} proposed to  solve for $f$, defined by the following problem
\begin{align} \label{Liouville-delta}
& \partial_t f + \nabla_\xi H \cdot \nabla_x f - \nabla_x H \cdot \nabla_\xi f=0, \\ \nonumber
& f(0, x,\xi)=\prod_{i=1}^d\delta(\xi_i - u_i(0,x)),
\end{align}
whose analytical solution is $f(t,x,\xi) = \delta(\phi(t,x,\xi))$. We have thus transformed a $(d+1)$-dimensional nonlinear Hamilton-Jacobi PDE  to a $(2d+1)$-dimensional \textit{linear} PDE~--~the Liouville equation, without \textit{any} approximations or constraints on the nonlinearity. The mapping is {\it exact}, but  at the expense of doubling the spatial dimension.

For geometric optics, the Liouville equation can be written as
\begin{equation}\label{Liouville-geo}
f_t + c(x) \frac{\xi}{|\xi|} \cdot \nabla_x f - |\xi|\nabla_x c \cdot  \nabla_{\xi} f = 0.
\end{equation}
The bicharacteristics of the Liouville equation \eqref{Liouville-geo} satisfy the Hamiltonian system:
\begin{equation}\label{bicharacteristics}
\frac{\d x}{\d t} = c(x) \frac{\xi}{|\xi|}, \qquad \frac{\d \xi}{\d t} = - |\xi|\nabla_x c.
\end{equation}
In particular, the 1-D Liouville equation is
\[f_t + c(x) \text{sign}(\xi) f_x - c_x |\xi| f_\xi = 0,\]
where $c(x)>0$ may be discontinuous at the interface between two media.

\subsubsection{The condition for transmissions and reflections at the interface}

\begin{figure}[!htb]
  \centering
  \includegraphics[scale=0.5]{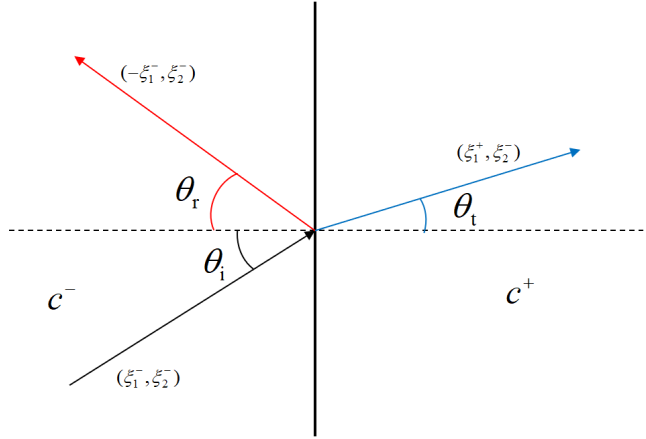}\\
  \caption{Wave transmission and reflection at an interface}\label{fig:interface}
\end{figure}

In geometrical optics, when a wave moves with its density distribution governed by the Liouville equation, its Hamiltonian $H = c|\xi|$ should be preserved across the interface
\begin{equation}\label{HPC}
H(x^+,\xi^+) = H(x^-,\xi^-) \qquad \mbox{or} \qquad c^-|\xi^-| = c^+ |\xi^+|,
\end{equation}
where the superscripts $\pm$ represent the right and left limits of the quantity at the interface. Let us consider a 2-D example. When a plane wave hits a flat vertical interface as shown in Fig.~\ref{fig:interface}, the Hamiltonian preserving condition \eqref{HPC} is equivalent to Snell's law of refraction
\[\frac{\sin \theta_{\text{i}}}{c^-} = \frac{\sin \theta_{\text{t}}}{c^+},\]
and the reflection law
\[\theta_{\text{r}} = \theta_{\text{i}},\]
where $\theta_{\text{i}}, \theta_{\text{t}}$ and $\theta_{\text{r}}$ stand for angles of incident, transmitted and reflected waves. Let $\xi = (\xi_1, \xi_2)$. Assume that the incident wave has a velocity $(\xi_1^-, \xi_2^-) $ to the left side of the interface, with $\xi_1^- > 0$. Since the interface is vertical ($\partial_y c = 0$), the characteristic of $\xi$ in \eqref{bicharacteristics} implies that $\xi_2$ is not changed when the wave crosses the interface. When $c^->c^+$, the wave can partially transmit and partially be reflected.  In this case, the local wave speed decreases, so the wave will cross the interface and increase its $\xi$ value in order to maintain a constant Hamiltonian. The preserving condition \eqref{HPC} implies
\[\xi_1^+ = \sqrt{\rho^2 (\xi_1^-)^2 + (\rho^2-1)(\xi_2^-)^2}, \qquad \rho = c^-/c^+.\]

As a linear hyperbolic equation, the solution to the Liouville equation \eqref{Liouville-geo}, can be obtained by the method
of characteristics. Namely, the density distribution $f$ remains a constant along a bicharacteristic. However, when partial transmissions and reflections are considered, this is no longer valid, since $f$ needs to be determined from two bicharacteristics, one accounting for the
transmission and the other for reflection. Ref.~\cite{JW06optics2} uses the following condition
at the interface:
\begin{equation}\label{interfaceSnell}
f(t,x^+,\xi^+) = \alpha_T f(t,x^-,\xi^-) + \alpha_R f(t,x^+, - \xi^+),
\end{equation}
where $\alpha_T, \alpha_R \in [0,1]$ are the transmission and reflection coefficients, satisfying $\alpha_T + \alpha_R = 1$, and $x^+ = x^-$ (for a sharp interface). Note that for a complete transmission, $f(t,x^+,\xi^+) = f(t,x^-,\xi^-)$,  while for a complete reflection, $f(t,x^+,\xi^+) = f(t,x^-,-\xi^-)$ and $\xi^- = \xi^+$.  For partial transmissions and reflections $\alpha_R, \alpha_T \in (0,1)$.

\subsection{The Hamiltonian-preserving scheme}

\subsubsection{The numerical flux}

We now describe the Hamiltonian-preserving finite difference scheme proposed in \cite{JW06optics1,JW06optics2} for the 1-D Liouville equation
\[f_t + c(x) \text{sign}(\xi) f_x - c_x |\xi| f_\xi = 0,\]
where $c(x)>0$ may be discontinuous.

We use a uniform mesh with grid points at $x_{i+\frac12}$, $i = 0,1,\cdots, N$ in the $x$-direction and $\xi_{j+\frac12}$,  $j = 0,1,\cdots, M$ in the $\xi$-direction.  The cells are centered at $(x_i,\xi_j)$ for $1\le i \le N$ and $1\le j \le M$, where $x_i = \frac12 (x_{i-\frac12} + x_{i+\frac12})$ and $\xi_j = \frac12(\xi_{j-\frac12} + \xi_{j+\frac12})$. The cell average of $f$ is defined by
\[f_{ij} = \frac{1}{\Delta x \Delta \xi}\int_{x_{i-\frac12}}^{x_{i+\frac12}} \int_{\xi_{j-\frac12}}^{\xi_{j+\frac12}} f(t,x,\xi) \d \xi \d x.\]

Assume that the discontinuous points of the wave speed $c$ are located at the grid points. Let the right and left limits of $c(x)$ at point $x_{i+\frac12}$ be $c_{j+\frac12}^+$ and $c_{i+\frac12}^-$, respectively.
We define the average wave speed as $c_i = \frac12 (c_{i-\frac12}^- + c_{i+\frac12}^+)$.
The flux splitting technique is adopted here. The semi-discrete scheme reads
\begin{equation}\label{semiODEsgeo}
\frac{\d}{\d t} f_{ij} + \frac{c_i \text{sign}(\xi_j)}{\Delta x}( f_{i+\frac12, j}^+ - f_{i-\frac12 ,j }^-)
- \frac{c_{i+\frac12}^+ - c_{i-\frac12}^-}{\Delta x \Delta \xi}|\xi_j| ( f_{i, j+\frac12} - f_{i, j-\frac12}) = 0
\end{equation}
for $1\le i \le N$ and $1\le j \le M$, where the numerical fluxes $f_{i, j+\frac12}$ in the $\xi$-direction are defined using the upwind discretisation, that is,
\[f_{i, j+\frac12} - f_{i, j-\frac12} = \begin{cases}
f_{ij} - f_{i,j-1}, \qquad & c_\xi \ge 0 \\
f_{i,j+1} - f_{i,j}, \qquad & c_\xi <0 \\
\end{cases} , \qquad c_\xi = - \frac{c_{i+\frac12}^+ - c_{i-\frac12}^-}{\Delta x \Delta \xi}|\xi_j|.\]

Since the characteristics of the Liouville equation may be different on the two sides of the interface, the corresponding numerical fluxes should also be different. The essential part of the algorithm is to define the split numerical fluxes $f_{i+\frac12,j}^{\pm}$ at the cell interface by utilizing the interface condition \eqref{interfaceSnell}.

Assume $c$ is discontinuous at $x_{i+ \frac12}$.
Consider the case $\xi_j > 0$. Since the wave moves from left to right in $[x_i, x_{i+\frac12}]$, we can define the interface value $f_{i+\frac12,j}^+ = f_{ij}$ using the upwind approximation. According to the interface condition \eqref{interfaceSnell},
\[f_{i+\frac12,j}^- = \alpha_T f(t,x_{i+\frac12}^+,\xi_j^+) + \alpha_R f(t,x_{i+\frac12}^-, - \xi_j^-), \]
where $\xi_j^+$ is obtained from $\xi_j^- = \xi_j$ from \eqref{HPC}.  Noting that $\xi_j^+$ may not be a grid point, we have to define it approximately. One can first locate the two cell centers that bound this velocity, and then use linear interpolation to evaluate the needed numerical flux at $\xi_j^+$.  The case $\xi_j<0$ can be treated similarly.

The algorithm of computing the numerical flux is summarized in Algorithm \ref{alg:flux}.
\begin{algorithm}[!htb]
\caption{Computation of the numerical flux in $x$-direction \label{alg:flux}}
\begin{enumerate}
  \item[Case 1:] $\xi_j > 0$.
  \begin{itemize}
    \item $f_{i+\frac12,j}^+ = f_{ij}$,  $\xi' = \frac{c_{i+\frac12}^-}{c_{i+\frac12}^+} \xi_j$.
    \item If $\xi_k \le \xi' < \xi_{k+1}$ for some $k$, then
\[a_R = \Big( \frac{c_{i+\frac12}^+ - c_{i+\frac12}^-}{c_{i+\frac12}^+ + c_{i+\frac12}^-} \Big)^2,  \qquad a_T = 1 - a_R,\]
\[f_{i+\frac12,j}^- = a_T \Big( \frac{\xi_{k+1}-\xi'}{\Delta \xi}f_{i,k} + \frac{\xi' - \xi_k}{\Delta \xi}f_{i,k+1} \Big)  + a_R f_{i+1, k'},\]
where $\xi_{k'} = - \xi_k$.
  \end{itemize}
  \item[Case 2:] $\xi_j < 0$.
   \begin{itemize}
    \item $f_{i+\frac12,j}^- = f_{i+1, j}$,  $\xi' = \frac{c_{i+\frac12}^+}{c_{i+\frac12}^-} \xi_j$.
    \item If $\xi_k \le \xi' < \xi_{k+1}$ for some $k$, then
\[a_R = \Big( \frac{c_{i+\frac12}^+ - c_{i+\frac12}^-}{c_{i+\frac12}^+ + c_{i+\frac12}^-} \Big)^2,  \qquad a_T = 1 - a_R,\]
\[f_{i+\frac12,j}^+ = a_T \Big( \frac{\xi_{k+1}-\xi'}{\Delta \xi}f_{i+1,k} + \frac{\xi' - \xi_k}{\Delta \xi}f_{i+1,k+1} \Big)  + a_R f_{i, k'},\]
where $\xi_{k'} = - \xi_k$.
  \end{itemize}
\end{enumerate}
\end{algorithm}

\subsubsection{The Schr\"odingerisation simulation}

We consider the first example in \cite{JW06optics2}. The discontinuous wave speed is given by
\[c(x) = \begin{cases}
0.6, \qquad & x<0,  \\
0.2, \qquad & x>0.
\end{cases}\]
The initial data is
\[f(0,x,\xi) = \begin{cases}
1, \qquad & x<0, \xi>0, \sqrt{x^2+4\xi^2} < 1, \\
1, \qquad & x>0, \xi<0, \sqrt{x^2+\xi^2} < 1, \\
0, \qquad & \mbox{otherwise}.
\end{cases}\]
The exact solution for $f$ at $t = 1$ is given by
\[
f(x, \xi, 1)
= \begin{cases}
\alpha^T, \qquad & 0<x<0.2, \quad \sqrt{1-(0.2-x)^2}<\xi<1.5 \sqrt{1-(3 x-0.6)^2}, \\
1 , \qquad & 0<x<0.2, \quad 0<\xi<\sqrt{1-(0.2-x)^2}, \\
1 , \qquad & 0<x<0.8, \quad-\sqrt{1-(x+0.2)^2}<\xi<0, \\
1 , \qquad &-0.4<x<0, \quad 0<\xi<\frac{1}{2} \sqrt{1-(x-0.6)^2}, \\
1 , \qquad & -0.6<x<0, \quad-\frac{1}{3} \sqrt{1-(\frac{x}{3}+0.2)^2}<\xi<0,\\
\alpha^R , \qquad & -0.6<x<0, \quad -\frac{1}{2} \sqrt{1-(x+0.6)^2}<\xi<-\frac{1}{3} \sqrt{1-(\frac{x}{3}+0.2)^2} ,\\
0  , \qquad &  \mbox{otherwise}.
\end{cases}
\]

\begin{figure}[!htb]
  \centering
  \subfigure[Numerical solution]{\includegraphics[scale=0.55]{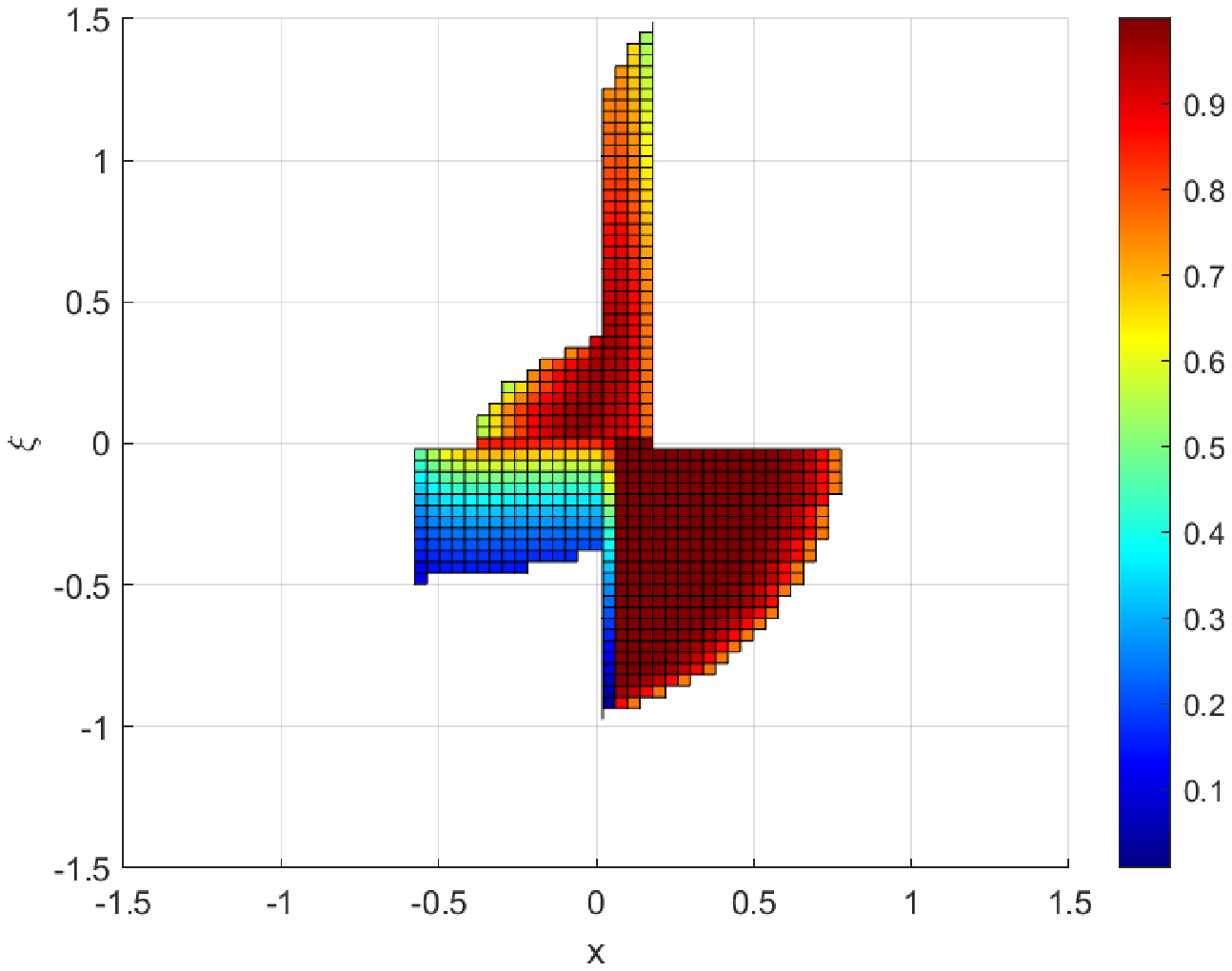}}
  \subfigure[Exact solution]{\includegraphics[scale=0.55]{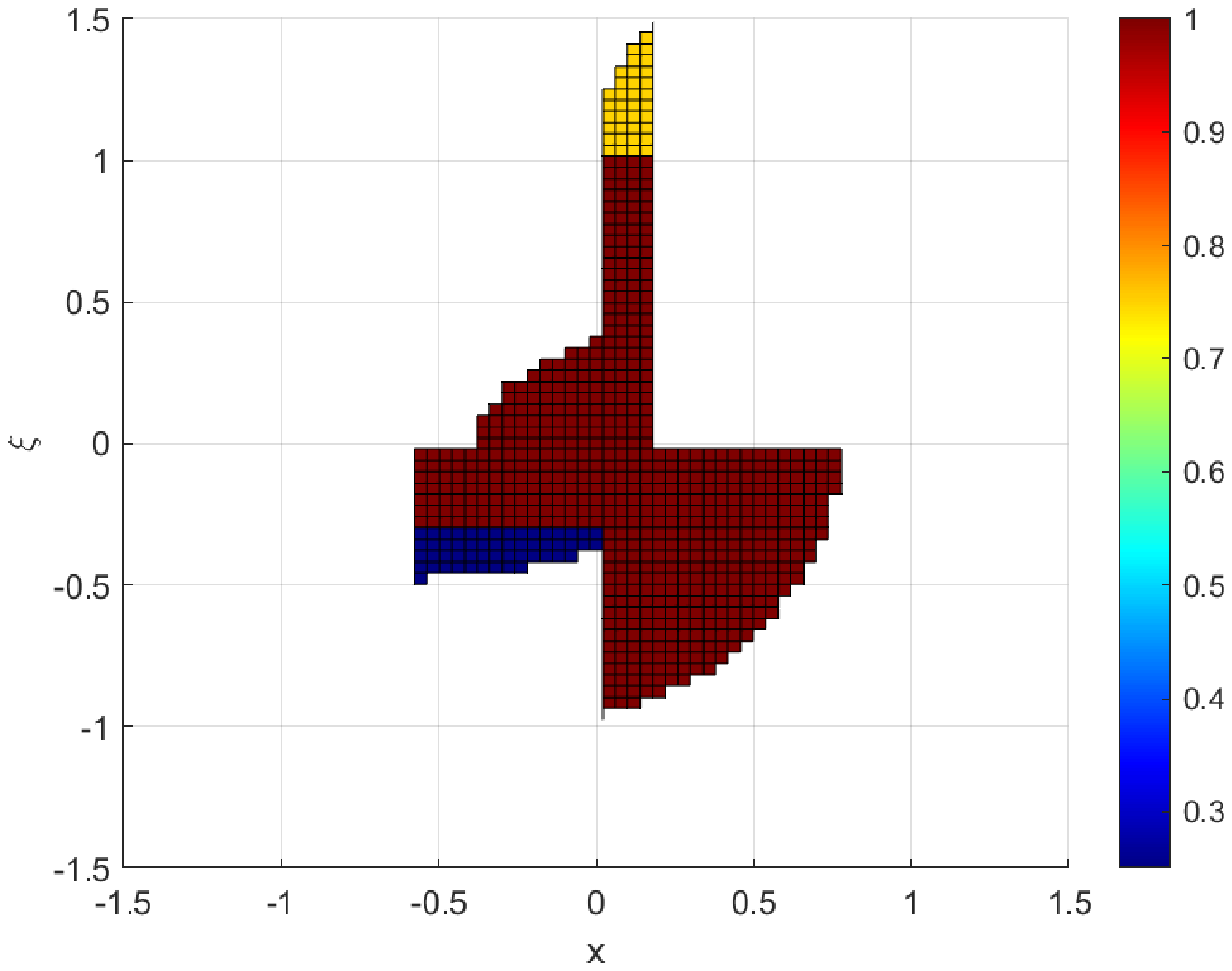}}\\
  \caption{Snapshots of the nonzero part of the solutions for the Hamiltonian preserving scheme.} \label{fig:geoHP}
\end{figure}

In the implementation, we choose a large enough domain that contains the supports of the initial and final solutions. For this example, one can take it as $[-1.5,1.5]^2$. To save the computational cost in the $p$-direction, where $p$ is the auxiliary variable for the Schr\"odingerisation approach, we set it as $[-4,4]^2$. For simplicity, we use the forward Euler method to iteratively get the updated solution. We take $N=M = 200$ and $N_t = 1000$, where $N_t$ is the number of steps for time discretisation. The numerical result is displayed in Fig.~\ref{fig:geoHP}. Due to the first-order accuracy in $x,\xi,p$ and $t$, the numerical solution has some smearing across the discontinuities, which is expected and can be improved by using more grid points or using higher order approximations.  We remark that for the direct upwind discretisation, the CFL condition requires the time step to satisfy $\Delta t = \mathcal{O}(\Delta x \Delta \xi)$. However, the Hamiltonian preserving scheme allows a time step $\Delta t = \mathcal{O}(\Delta x, \Delta \xi)$.

\section{Conclusion}

Quantum simulations for time-dependent or independent boundary value problems of partial differential equations are quite difficult because the ODE system resulting from spatial discretisations is not necessarily a Hamiltonian system. Spatial discretisation of the boundary condition, like the Dirichlet boundary condition for example, could also give rise to an inhomogeneous term in the system ( $\bb{b} \neq \bb{0}$ in \eqref{ODElinear}). Our Schr\"odingerisation approach combined with the augmentation technique resolves this problem in a generic and efficient way as shown in \cite{JLLY23ABC} and this work.

In this article, we extend the Schr\"odingerisation approach \cite{JLY22SchrShort, JLY22SchrLong} for quantum simulations of PDEs to problems with physical boundary or interface conditions. While a quantum dynamics with physical boundary or interface conditions is no longer a Hermitian Hamiltonian system, the Schr\"odingerisation approach makes it so in a simple fashion.
We give the implementation details for these problems. The numerical experiments validate this approach, demonstrating that the Schr\"odingerised systems yield the same results as the original dynamics. This further extends the Schr\"odingeration techniques toward real applications of partial differential equations which most often are coupled with boundary or interface conditions.

\section*{Acknowledgements}
SJ was partially supported by the NSFC grant No.~12031013, the Shanghai Municipal Science and Technology Major Project (2021SHZDZX0102), and the Innovation Program of Shanghai Municipal Education Commission (No. 2021-01-07-00-02-E00087).  NL acknowledges funding from the Science and Technology Program of Shanghai, China (21JC1402900). YY was partially supported by China Postdoctoral Science Foundation (no. 2022M712080). XL is supported by a Seed Grant from the Institute of Computational and Data Science (ICDS) at Penn State.

\bibliographystyle{alpha} 

\newcommand{\etalchar}[1]{$^{#1}$}

\end{document}